\documentclass{appolb}
\usepackage{epsfig}
\usepackage{graphics}
\usepackage{amssymb,amsmath}
\usepackage{graphicx}
\usepackage{color}
\usepackage{todonotes}

\begin{document}
\title{Inhomogeneous islands and continents in the 
Nambu--Jona-Lasinio model%
\thanks{Presented at the HIC for FAIR Workshop and XXVIII Max Born Symposium \emph{Three days on Quarkyonic Island}, Wroc\l aw, 19-21 May, 2011}%
}
\author{Stefano Carignano
\and
Michael Buballa
\address{Institut f\"ur Kernphysik, Technische Universit\"at Darmstadt, Germany}
}
\maketitle
\begin{abstract}
We present some recent developments in our study of inhomogeneous chiral 
symmetry breaking phases in the Nambu--Jona-Lasinio model.
First, we investigate different kinds of one- and two-dimensional
spatial modulations of the chiral condensate within the inhomogeneous 
``island'' and compare their free energies.
Next, we employ the Polyakov-loop extended version of the model to study
the effects of varying the number of colors on the inhomogeneous region.
Finally, we discuss the properties of an inhomogeneous ``continent'' 
which appears in our model at higher chemical potentials, 
and analyze its origin.
\end{abstract}
\PACS{21.65.Qr,12.38.Mh,12.39.Fe}

\section{Introduction}

The study of spatially modulated ground states in strongly interacting systems 
is an old topic \cite{Overhauser:1962,Goldhaber:1987,Dautry:1979,Broniowski:1990dy} 
(see Ref.~\cite{Broniowski:2011} for a recent review)
which has recently received new attention.
It has been argued some time ago that the favored ground state of a dense
Fermi sea of quarks should be characterized by a spatial modulation 
of the chiral condensate, at least in the limit of a large number of colors 
($N_c$)~\cite{Deryagin:1992,Shuster:1999}.
More recent studies, especially in the context of quarkyonic matter,
seem to support this hypothesis~\cite{Kojo:2009ha,Kojo:2010,Kojo:2011}.

But also for physical case of three colors, 
Nambu--Jona-Lasinio-- (NJL--) type model
studies have revealed the presence of an inhomogeneous ``island'' 
at intermediate chemical potentials and low temperatures, 
namely in the region where the usual first-order chiral phase transition
would occur when limiting to homogeneous phases 
only~\cite{Sadzikowski:2000,NT:2004,Nickel:2009}.
In particular, the critical endpoint of the phase boundary is covered
by the inhomogeneous region and thus disappears from the phase
diagram~\cite{CNB:2010}.

In these proceedings we present some of the most recent results in 
our NJL-model study of inhomogeneous chiral symmetry breaking phases. 
After briefly outlining our methods for tackling this problem 
and the technical difficulties associated with it, we show numerical 
results for different kinds of one- and two-dimensional modulations and 
compare their free energies.
In order to build a bridge towards large-$N_c$ studies, we then consider 
a Polyakov-loop extended NJL model for an arbitrary (large) number of colors 
and see how this modification alters the inhomogeneous phase. 
Finally, we discuss the properties of a second inhomogeneous region, 
which appears in our model at higher chemical potentials, 
and try to understand the origin of this new ``continent''. 
In particular, we investigate whether it
is to be interpreted as a regularization artifact 
or as a simple consequence of the model interaction.

\section{Inhomogeneous phases in NJL}
Starting from the two-flavor Nambu-Jona Lasinio Lagrangian \cite{NJL}, 

\begin{equation}
\mathcal{L}_{NJL} = \bar\psi (i\gamma^\mu\partial_\mu -m)\psi + G \left( ({\bar\psi\psi})^2 + (\bar\psi i \gamma^5 \tau_a \psi)^2\right)\,,
\end{equation}
we perform the mean-field approximation, allowing for a spatial dependence of the scalar and pseudoscalar condensates:

\begin{equation}
\langle\bar{\psi}\psi\rangle=S(\vec x)  \,,\qquad 
\langle\bar{\psi}i\gamma^5\tau^a\psi\rangle=P(\vec x)\,\delta_{a3} \,.
\end{equation}
The mean-field Lagrangian can be rewritten by introducing an effective Hamilton operator $\mathcal{H}$ \cite{Nickel:2009,NB:2009}:

\begin{equation}
\mathcal{L_{MF}} = 
\bar\psi \gamma^0 (i\partial_0 - \mathcal{H})\psi - G (S^2 + P^2) \,,
\label{eq:LMF}
\end{equation}
with $\mathcal{H}$ defined as

\begin{equation}
\mathcal{H} =  \gamma^0 \left[ i\vec\gamma\cdot\vec\partial + m - 2G(S + i \gamma^5 \tau_a P) \right] \,.
\end{equation}
Explicitly, we can write down $\mathcal{H}$ as \cite{Nickel:2009}

\begin{equation}
\mathcal{H} = 
\left( 
\begin{array}{cc}
 -i\vec\sigma\cdot\vec\partial &  M(\vec x) \\
 M^*(\vec x) & i\vec\sigma\cdot\vec\partial \\
\end{array} 
\right) \,,
\end{equation}
where the chiral representation for the Dirac matrices has been used
and we introduced an effective inhomogeneous ``mass'' function

\begin{equation}
M(\vec x) = m - 2G (S(\vec x) + i  P(\vec x)) \,.
\end{equation}
In order to evaluate the thermodynamic potential associated to this (so far generic) spatially modulated chiral condensate, we work, as customary, in 
imaginary time and switch to momentum space. By assuming static (\ie time-independent) condensates, we are able to perform explicitly the 
sum over Matsubara frequencies and obtain the expression (up to a constant)

\begin{equation}
\label{eq:Omega2}
\Omega(T,\mu;M(\vec x))
=
\Omega_\mathit{kin}(T,\mu;M(\vec x))
+
\Omega_\mathit{cond}(M(\vec x))
\,,
\end{equation}
with
\begin{equation}
\label{eq:Omegacond}
\Omega_\mathit{cond}(M(\vec x))
=
\frac{1}{V} \int_V\, d^3x \,\frac{\vert M(\vec x)-m\vert^2}{4G}
\,,
\end{equation}
where $V$ is the volume of the system, and
\begin{equation}
\label{eq:Omegakin}
\Omega_\mathit{kin}(T,\mu;M(\vec x))
=
-T
\sum_{E_{n}}
\log\left(2\cosh\left(\frac{E_{n}-\mu}{2T}\right)\right)
\,,
\end{equation}
where the sum is over all eigenvalues $E_n$ of $\mathcal{H}$ in color, 
flavor, Dirac and momentum space.

For homogeneous condensates the diagonalization of~$\mathcal{H}$ in Dirac 
space simply gives twice $E_p = \pm\sqrt{p^2+M^2}$ and the eigenvalue sum may 
be trivially turned into an integral over all possible momenta.
In presence of an inhomogeneous condensate, however, the diagonalization of 
$\mathcal{H}$ is a highly non-trivial task,
since the quarks may exchange momenta by scattering off the inhomogeneous 
condensate. This means that the spatially modulated mass term will effectively 
couple quarks with different momenta and the resulting structure will not be 
diagonal in momentum space.

In the following we assume to have a lattice structure and, thus, a periodic
shape of the modulation. This means that we can expand the modulated chiral
condensate in a Fourier series,
\begin{equation}
M(\vec x) = \sum_{\vec{q}_k} M_{\vec{q}_k} e^{i \vec{q}_k\cdot\vec x} \,,
\label{eq:Mq}
\end{equation}
with discrete momenta $\vec{q}_k$, which form a reciprocal lattice (RL). 
A generic element of $\mathcal{H}$ in momentum space then takes the form

\begin{equation}
\mathcal{H}_{\vec{p}_m,\vec{p}_n} =
 \left( 
\begin{array}{cc}
 -\vec\sigma\cdot\vec{p}_m\,\delta_{\vec{p}_m,\vec{p}_n} &  
 \sum_{\vec{q}_k} M_{\vec{q}_k} \delta_{\vec{p}_n,\vec{p}_m+\vec{q}_k} 
 \\
 \sum_{\vec{q}_k} M^*_{\vec{q}_k} \delta_{\vec{p}_n,\vec{p}_m-\vec{q}_k} &  
 \vec\sigma\cdot\vec{p}_m\,\delta_{\vec{p}_m,\vec{p}_n} 
\end{array} 
\right) \,, 
\label{eq:mfH}
\end{equation}
where $\sum_{\vec{q}_k}$ runs over all momenta of the RL, making obvious
the non-diagonal structure of the matrix.
In turn, momenta which do not differ by an element of the RL are not
coupled, so that $\mathcal{H}$ can be decomposed into a block diagonal
form, where each block can be labelled by a momentum of the first
Brillouin zone (BZ).
This allows to decompose the eigenvalue sum in Eq.~(\ref{eq:Omega2}) into 
a momentum integral over the BZ times a sum over the discrete eigenvalues 
of each block.

For the simplest case of the so-called chiral density wave (CDW),
namely a plane wave characterized by a single momentum\footnote{
Some authors introduce an additional factor of two in the exponent,
$M(\vec x) = M e^{i 2\vec{q}\cdot\vec{x}}$.
This is motivated by the fact that the favored value of $|\vec Q|$
is roughly of the order of $2\mu$, so that $|\vec q|$ is of the order
of $\mu$.
Here we prefer the definition without the factor of two, which is 
more consistent with the general ansatz of Eq.~(\ref{eq:Mq}) and other
modulations studied in this article.}

\begin{equation}
M(\vec x) = M e^{i \vec{Q}\cdot\vec{x}} \,,
\label{eq:spiral}
\end{equation}
the sum in (\ref{eq:mfH}) is then obviously given by a single term, and the resulting BZ-projected matrix within this framework 
is characterized by a narrow band structure, with only 
the first off-diagonal block filled with nonzero entries.

For general periodic structures, 
although the numerical diagonalization procedure is in principle straightforward, its practical implementation turns out to be computationally demanding.
Therefore, in order to simplify the problem, we limit the generality 
of our ansatz Eq.~(\ref{eq:Mq}) for the spatially modulated condensate
and consider simpler, lower-dimensional modulations.

\section{One-dimensional modulations}
The eigenvalue problem simplifies considerably when the chiral condensate is allowed to vary only in one spatial dimension.
In this special case it has been observed that the dimensionally reduced effective Hamiltonian becomes formally identical to that of the 1+1-dimensional Gross-Neveu model \cite{Nickel:2009},
for which self-consistent solutions are already well known \cite{Schnetz:2004vr,Schnetz:2005ih,Thies:2006ti,Basar:2009fg}. One finds  
(from now on we will consider only results in the chiral limit for simplicity)

\begin{equation}
M(z) = \sqrt{\nu}\Delta sn(\Delta z\vert\nu) \,,
\label{eq:msolitons}
\end{equation}
where $sn(\Delta z\vert\nu)$ is a Jacobi elliptic function \cite{abramovitz}.
For this kind of solutions, an analytical expression for the eigenvalue spectrum, depending on the elliptic parameter $\nu$ and the amplitude of the condensate $\Delta$ can be obtained 
and the minimization of the thermodynamic potential may easily be performed with respect to these two quantities \cite{Nickel:2009}.
No numerical diagonalization of $\mathcal{H}$ is thus needed.
In particular, Eq.~(\ref{eq:Omegakin}) can be written as

\begin{equation}
\Omega_\mathit{kin}
= 
-N_fN_c\int_0^\infty \vspace{-2mm}dE\,\rho(E) 
(f_\mathit{vac} + f_\mathit{med})\,, 
\label{eq:omegarho}
\end{equation}
where all functions which enter the integrand are known 
analytically~\cite{Nickel:2009}: 
$\rho(E)$ is the spectral density associated to this kind of solutions,
$f_{vac}$ is the (divergent) vacuum contribution related to the Dirac sea, 
which we regularize by introducing Pauli-Villars-type 
counterterms~\cite{Klevansky:1992qe},

\begin{equation}
f_{vac}(E,\Lambda) = \sum_{j=0}^3 c_j \sqrt{E^2+j\Lambda^2}\,, 
\qquad
c_0 = -c_3 =1,\; c_2 = -c_1 = 3\,, 
\label{eq:fuv}
\end{equation}
and $f_{med}$ describes the medium part:

\begin{equation}
f_{med}(E) = T \log \left(1 + e^{-(E - \mu)/T}\right) +
             T \log \left(1 + e^{-(E + \mu)/T}\right)
\,.
\label{eq:fmed}
\end{equation}

\begin{figure}
\begin{center}
\includegraphics[width=.45\textwidth]{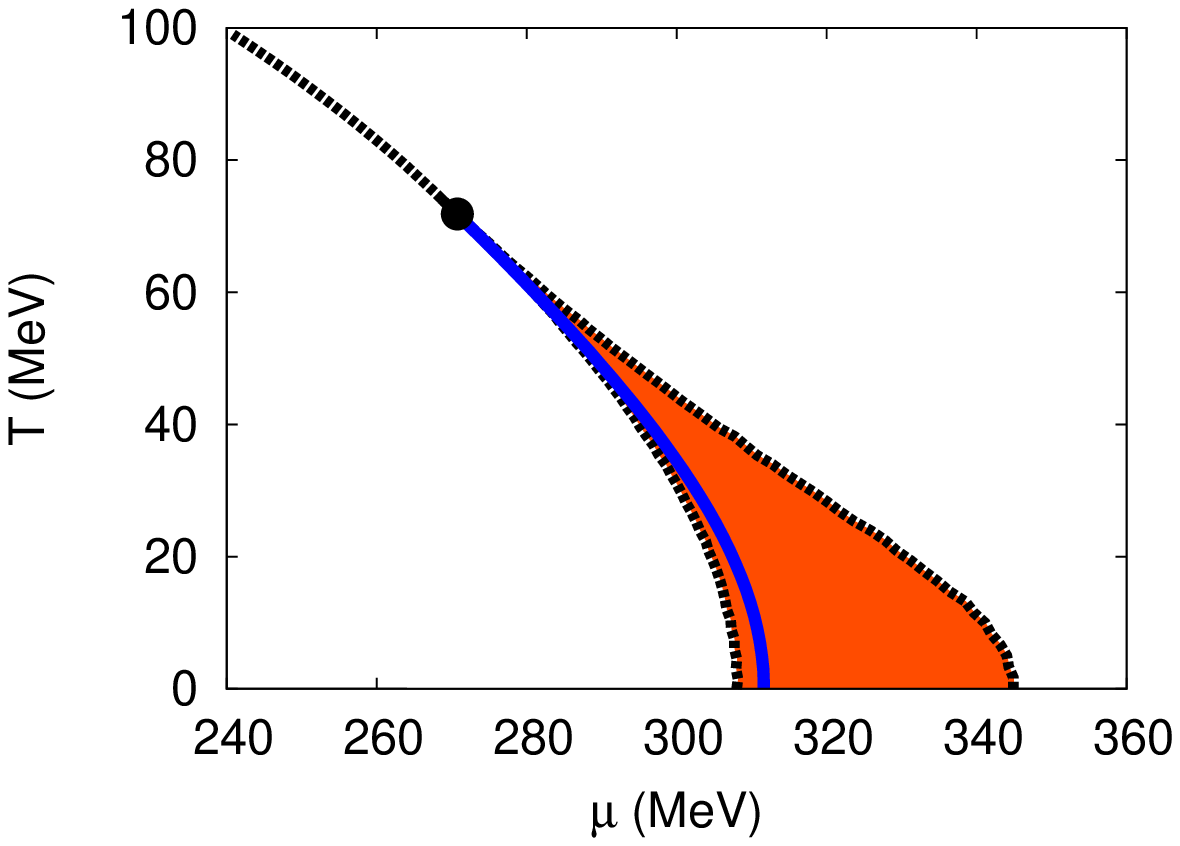}
\includegraphics[width=.45\textwidth]{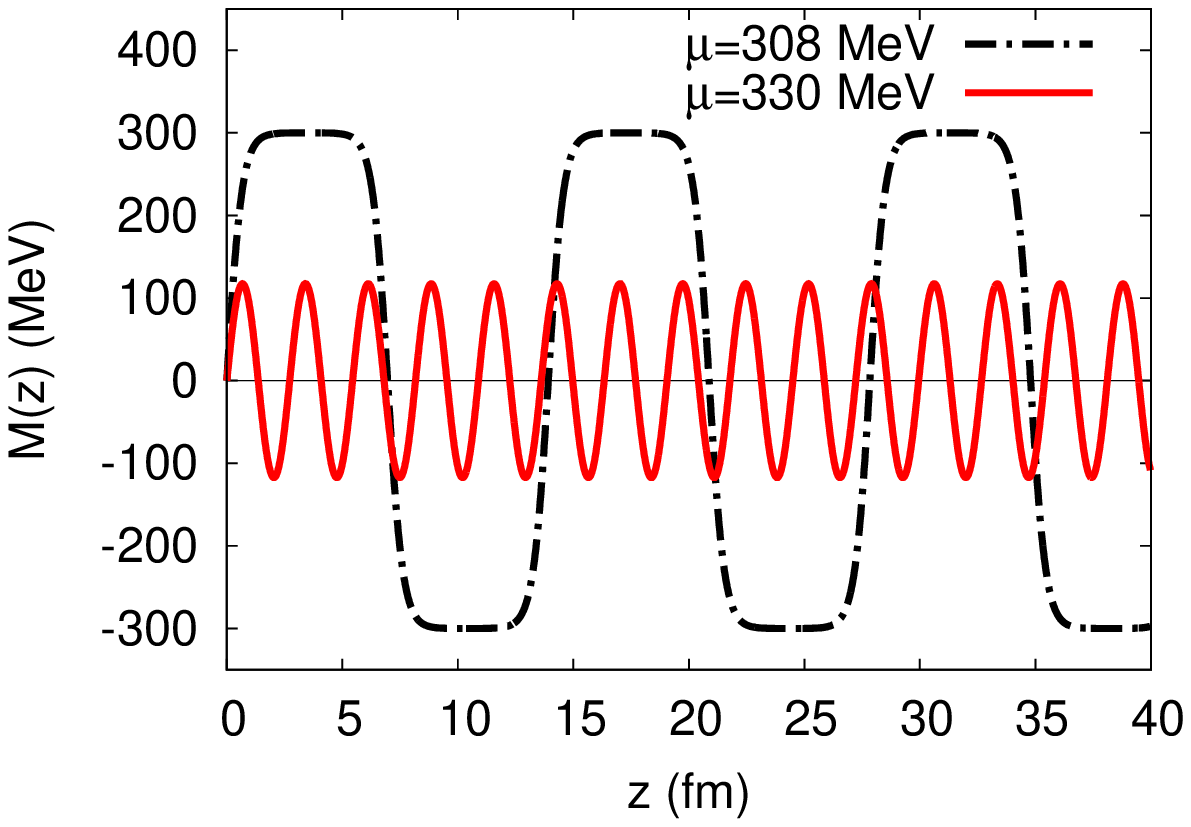}
\end{center}
\caption{Left:
Phase diagram obtained when allowing for solitonic solutions.
The black dotted lines represent the second-order transition lines 
separating the homogeneous chirally broken phase (left), the 
chirally restored phase (right) and the inhomogeneous phase (shaded
region).
The blue solid line represents the first-order phase boundary
obtained when limiting to homogeneous order parameters.
It is completely covered by the inhomogeneous phase and ends at
the ``Lifshitz point'' where the three second-order phase boundaries
meet.
Right: Mass function $M(z)$ at $T=0$ for two different values of the 
chemical potential.
} 
\label{fig:pdsolitons}
\end{figure}

The resulting phase diagram is shown in Fig.~\ref{fig:pdsolitons}.
It features a region at low temperatures and intermediate chemical potentials 
where an inhomogeneous phase is favored over the homogeneous chirally broken 
and restored solutions. 
At the onset of this inhomogeneous island the chiral condensate assumes 
the shape of a single soliton 
(which is thermodynamically degenerate with the homogeneous broken solution) 
and then progressively varies into 
a more sinusoidal shape until it gradually melts when reaching the restored 
phase. In this case, all phase transitions 
are second order \cite{Nickel:2009,CNB:2010}.

\begin{figure}
\begin{center}
\includegraphics[width=.45\textwidth]{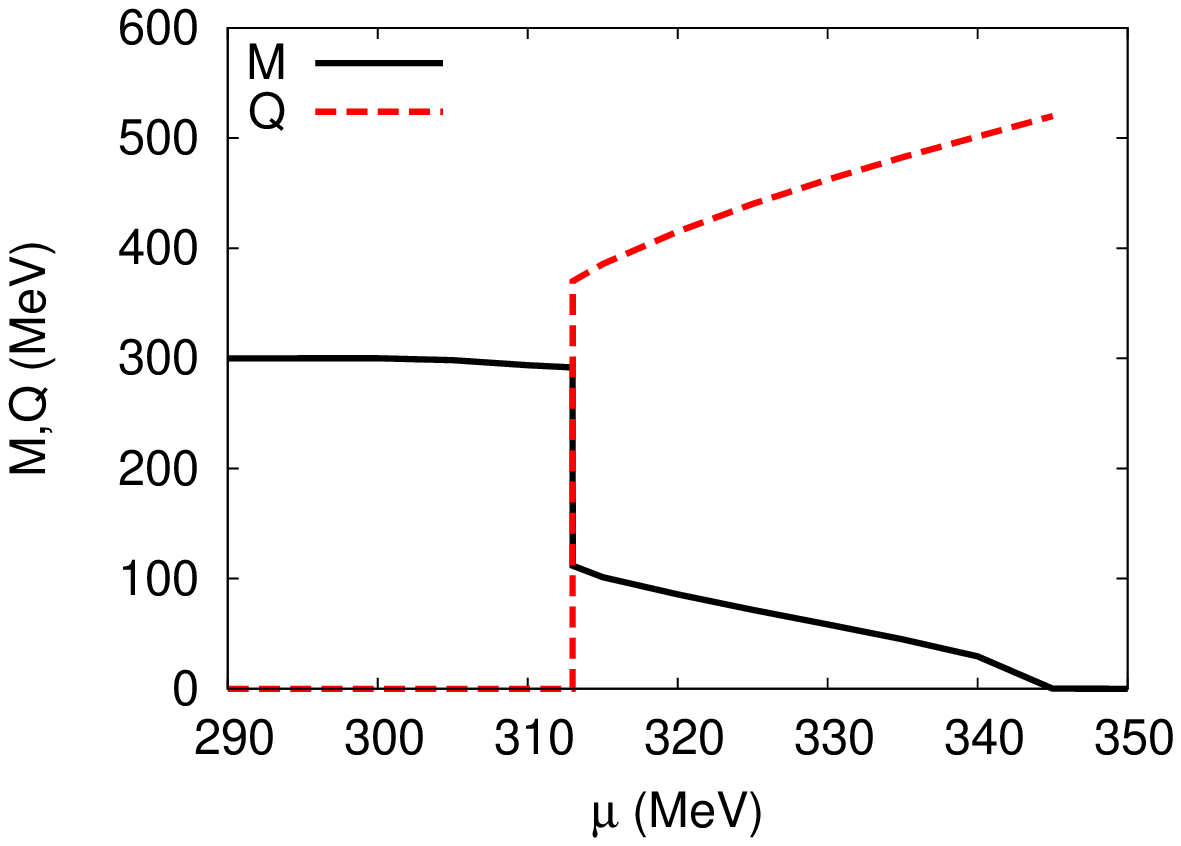}
\includegraphics[width=.45\textwidth]{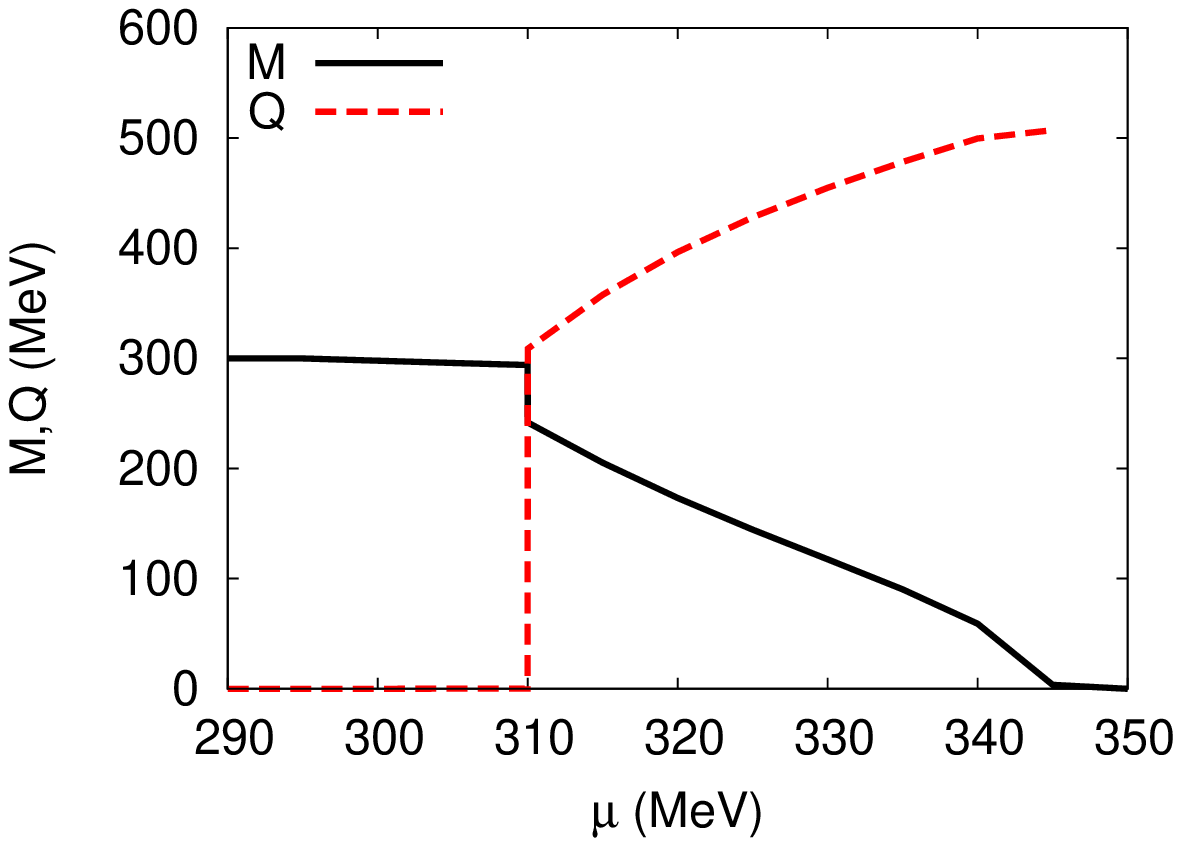}
\end{center}
\caption{Amplitude and wave number at $T=0$ as functions of the chemical
potential for the CDW ansatz, Eq.~(\ref{eq:spiral}) (left) 
and for sinusoidal modulations, Eq.~(\ref{eq:cos}) (right).} 
\label{fig:cosspiral}
\end{figure}

It is of course possible to restrict the one-dimensional ansatz to something even simpler, 
such as a chiral density wave (\ref{eq:spiral}) \cite{Sadzikowski:2000,NT:2004} or a real sinusoidal modulation,
\begin{equation}
M(z) = M cos(Q z) \,.
\label{eq:cos}
\end{equation}
When implementing the CDW modulation one finds that the BZ-projected $\mathcal{H}$ is immediately block-diagonal and an analytical expression for the density
of states can be obtained straightforwardly. The thermodynamic potential 
can then again be worked out
using Eq.~(\ref{eq:omegarho}), with the only difference
being a different expression for $\rho(E)$ \cite{Nickel:2009}. 
For the real sinusoidal modulation, on the other hand,  
a brute-force numerical diagonalization in Dirac and momentum space is required.
Results for the order parameters of these two kinds of modulation are shown in Fig.~\ref{fig:cosspiral}. 
After comparing the free energy of these modulations 
(see also section \ref{sec:freeen}), 
one finds that both are favored over the homogeneous 
solutions in a very similar window in chemical potential. 
In fact, it has been argued \cite{Nickel:2009prl} that all these inhomogeneous 
solutions should share a common second order transition 
to the restored phase, while the onset from the chirally broken phase need not be the same, since in this case the two phases are separated by a first order line\footnote{While the solitonic ansatz
may assume a shape that is thermodynamically degenerate with the homogeneous broken case, there is no smooth way for these less general solutions to approach a spatially constant shape
unless for the trivial case $M = 0$.}.
Nevertheless,
numerical results show how the onsets of the two phases occur at roughly the same value of chemical potential, with a slightly larger window for the real sinusoidal modulation.

\section{2D modulations}
While the study of one-dimensional modulations has been able to provide a 
first insight on the importance of 
spatial modulations of the chiral condensate, 
it is natural to expect that also higher-dimensional modulations could
appear in the phase diagram of a 3+1 dimensional system.
Aside from being a more general ansatz, higher-dimensional modulations are also
unaffected by the instability with respect to fluctuations which prevents
the formation of a true 1D crystalline structure at finite temperature \cite{Baym:1982}.

The procedure for handling a higher-dimensional modulation is in principle 
identical to that already outlined for the one-dimensional case. 
However, for higher dimensions
no straightforward analytical results are available 
and the numerical diagonalization involves a bigger matrix, 
since one is dealing with two- or three-dimensional momenta in $\mathcal{H}$. 

An obvious next step is to study two-dimensional modulations.
Then the thermodynamic potential takes the form 

\begin{equation}
\Omega
= -N_f N_c\int_{-\infty}^\infty \frac{dp_\perp}{2\pi}\int_{BZ} \frac{d^2k}{(2\pi)^2} \sum_{\lambda_{\vec k}} T \log \left[ 2\cosh \left(\frac{E_\perp - \mu}{2T}\right) \right] 
+ \Omega_\mathit{cond}
\,,
\label{eq:omega2d}
\end{equation}
where $p_\perp$ is the momentum perpendicular to the directions of the spatial modulation 
and one introduces $E_\perp = sgn(\lambda_{\vec k})\sqrt{\lambda_{\vec k}^2 + p_\perp^2}$, with $\lambda_{\vec k}$ being the eigenvalues of the dimensionally-reduced $\mathcal{H}$, depending on the momentum $\vec k \in BZ$.

\subsection{Square modulation: egg carton}
\begin{figure}
\begin{center}
\includegraphics[width=.45\textwidth]{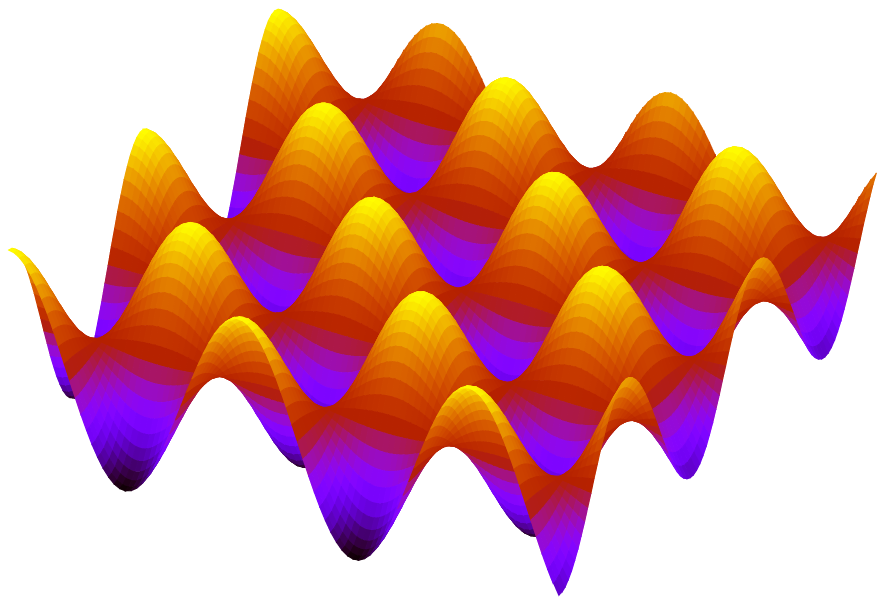}
\includegraphics[width=.45\textwidth]{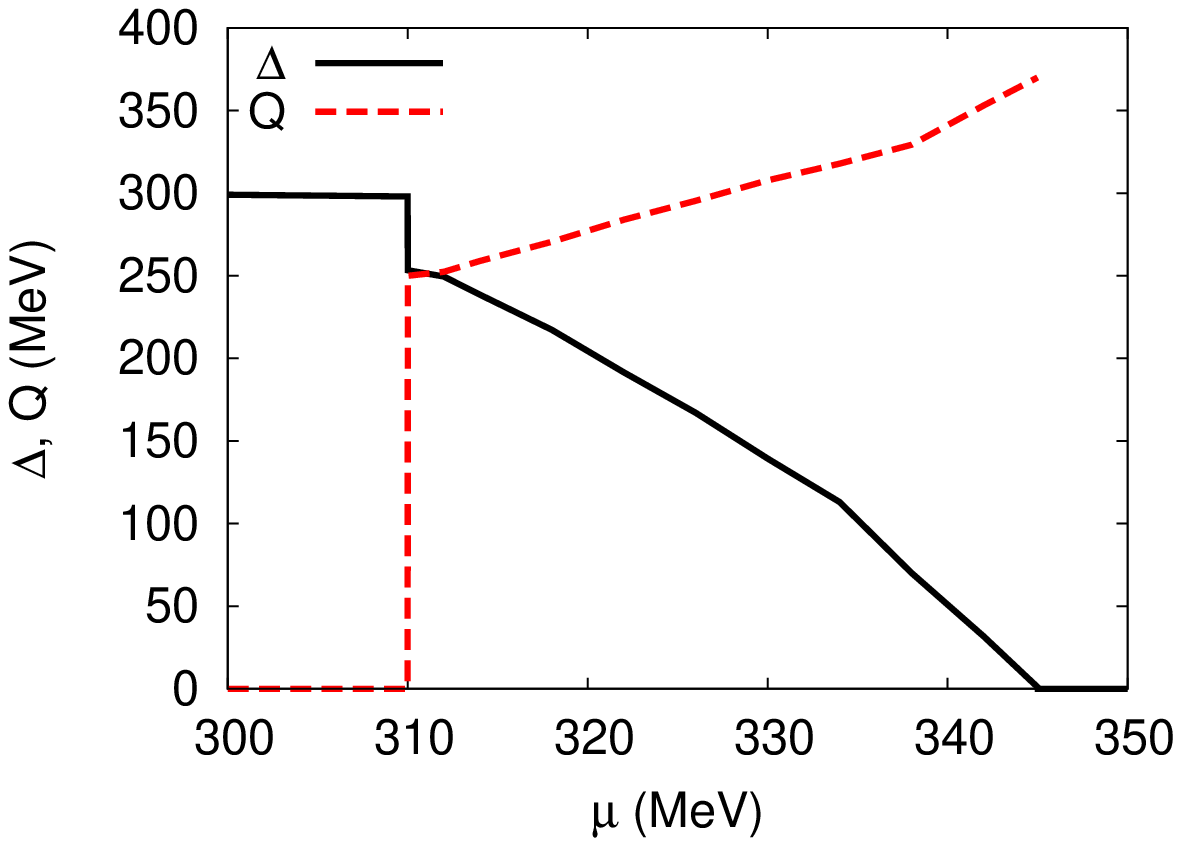}
\end{center}
\caption{Left: Shape of the egg-carton modulation. 
Right: Results for $Q$ and $\Delta$ obtained by numerical minimization of the 
thermodynamic potential for this two-dimensional ansatz.}
\label{fig:cos2d}
\end{figure}

Since there are no known results to guide the choice for the shape of the 
spatial modulation, we should in principle study different geometries 
of the two-dimensional crystal and for each of them 
minimize the thermodynamic potential with respect to the Fourier components 
$M_{\vec q}$ on the corresponding RL.
As a first step in this direction we consider a square lattice 
and a simple product of cosines,
\begin{equation}
M(x,y) = \Delta \cos(Q x)\cos(Q y) \,,
\end{equation}
thus restricting ourselves to the first harmonics only in each direction.
This highly symmetric ``egg carton'' (shown in Fig.~\ref{fig:cos2d}) 
is possibly the simplest real modulation one could think of in 
two dimensions.\footnote{This ansatz is equivalent to a modulation 
of the form $M(x',y') = \Delta'[\cos(Q' x')+ \cos(Q'y')]$,
which is one of the shapes discussed in \cite{Kojo:2011}.
Here $\Delta' = \Delta/2$,
$Q' = \sqrt{2} Q$ and $(x',y')$ is related to $(x,y)$ by a 
rotation of $\pi/4$ about the $z$-axis.}

After diagonalizing $\mathcal{H}$ numerically, the thermodynamic potential (\ref{eq:omega2d}) is minimized with respect 
to the two parameters $\Delta$ and $Q$, characterizing amplitude and period of the modulation, respectively.
The results of the numerical minimization are presented on the right-hand 
side of Fig.~\ref{fig:cos2d}.
Their main features are in qualitative agreement with the one-dimensional
examples discussed before.
We find again a sharp onset around $\mu \approx 310$ MeV and a smooth 
approach to the restored phase, which is reached at the same chemical 
potential as for the 1D modulations via a second-order phase transition.

\subsection{Comparison of different crystalline phases}
\label{sec:freeen}

The results presented in Figs.~\ref{fig:cosspiral} and \ref{fig:cos2d}
have been obtained by assuming a single ansatz for the mass 
modulation (CDW, cosine or egg carton) in each case.
Under this restriction we found that the different inhomogeneous phases  
are energetically favored over the homogeneous solutions in a window, 
which at $T=0$ and for our model parameters lies roughly between 
$\mu \approx 300$ and $350$ MeV.

The next obvious step is now to compare the values of the thermodynamic 
potential of these solutions with each other in order to find out
which of them has the lowest free energy.
In fact, for the one-dimensional modulations, we know already that the CDW 
and the cosine are disfavored against the solitonic solutions, 
but there could still be higher-dimensional modulations, which have
an even lower free energy than the latter.
In particular, it has been argued in the context of quarkyonic matter studies
that with increasing density such higher-dimensional solutions will be 
favored, at least in the limit of a large number of 
colors~ \cite{Kojo:2010,Kojo:2011}.  
On the other hand, using Ginzburg-Landau (GL) arguments, 
it has been argued that close to the Lifshitz point 1D modulations 
should be favored over higher dimensional ones within this type of 
models~\cite{Abuki:2011}.
We therefore focus on what happens for $T=0$, where a GL analysis is unable
to give reliable results.

\begin{figure}
\begin{center}
\includegraphics[width=.6\textwidth]{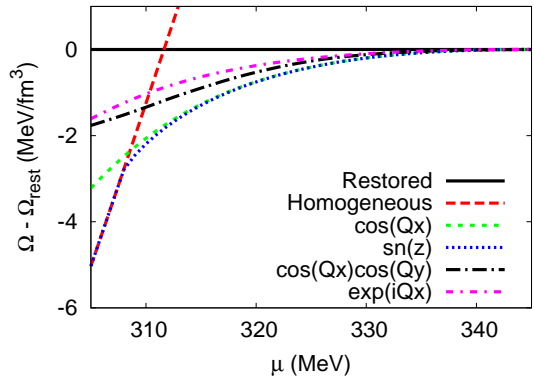}
\end{center}
\caption{Comparison of thermodynamic potentials relative to the restored
phase for different kinds of modulations at $T=0$.} 
\label{fig:cfrtp}
\end{figure}

The results of this comparison are shown in Fig.~\ref{fig:cfrtp}.
One can clearly see that the solitonic solutions (\ref{eq:msolitons}) lead 
to the biggest gain in free energy compared to all the other cases considered. 
The two-dimensional ``egg-carton'', on the other hand, turns out to be 
energetically disfavored with respect to one-dimensional real modulations 
throughout the whole inhomogeneous window, while still being favored over 
the chiral density wave ansatz (\ref{eq:spiral}). 

While a thorough discussion will require the study of several kinds of 
2D modulations, these preliminary results seem to indicate that the gain in 
condensation energy 
is not enough to compensate the greater kinetic energy cost due to the
presence of a condensate varying in more than one spatial dimensions, and 
thus 1D modulations remain the most favored.

 \section{PNJL and large $N_c$}
In order to mimic features of confinement, in particular to suppress the contribution of free constituent
quarks in the confined phase and to include gluonic contributions to the pressure,
the NJL model can be coupled to an effective description of the Polyakov loop \cite{Meisinger:1995ih,Fukushima:2003fw,Megias:2004hj,Ratti:2005jh}.
To this end, the quarks are minimally coupled to a
background gauge field.
Furthermore, a local potential $\mathcal{U}$,
which is essentially constructed to reproduce ab-initio results of
pure Yang-Mills theory at finite temperature,
is added to the thermodynamic potential.
The resulting model
is known as the PNJL model, and within the mean-field treatment 
the traced expectation value of the Polyakov loop, $\ell$,
becomes a new quantity which has to be determined self-consistently.

This extension is also possible when dealing with inhomogeneous 
phases and has been investigated in Ref.~\cite{CNB:2010} under the simplifying
assumption that the Polyakov loop expectation value $\ell$ is spatially 
uniform.
It was found that the effects of coupling
to the Polyakov loop basically amount to stretching the phase diagram 
towards higher temperatures, since the effective confinement
suppresses the thermal excitation of single quarks.

On the other hand, the PNJL model has also been used to investigate the 
influence of the number of colors on the phase 
diagram~\cite{McLerran:2008ua}. 
Here the coupling to the Polyakov loop is crucial, since the 
NJL mean-field results without Polyakov loop are $N_c$-independent.
The study of Ref.~\cite{McLerran:2008ua} was restricted to homogeneous 
phases, and its main focus was to see how the region of confined but 
chirally restored matter develops.
Some time ago, this was thought to be a possible manifestation of 
quarkyonic matter, whereas according to the present picture
chiral symmetry is inhomogeneously broken in the quarkyonic 
phase~\cite{Kojo:2009ha,Kojo:2010,Kojo:2011}.
Therefore, since the latter has been derived in the large-$N_c$ limit,
it is particularly interesting to combine the approaches of 
Refs.~\cite{CNB:2010} and \cite{McLerran:2008ua}, and to study the 
behavior of the inhomogeneous phase in the PNJL model at large 
$N_c$.

\begin{figure}
\begin{center}
\includegraphics[width=.32\textwidth]{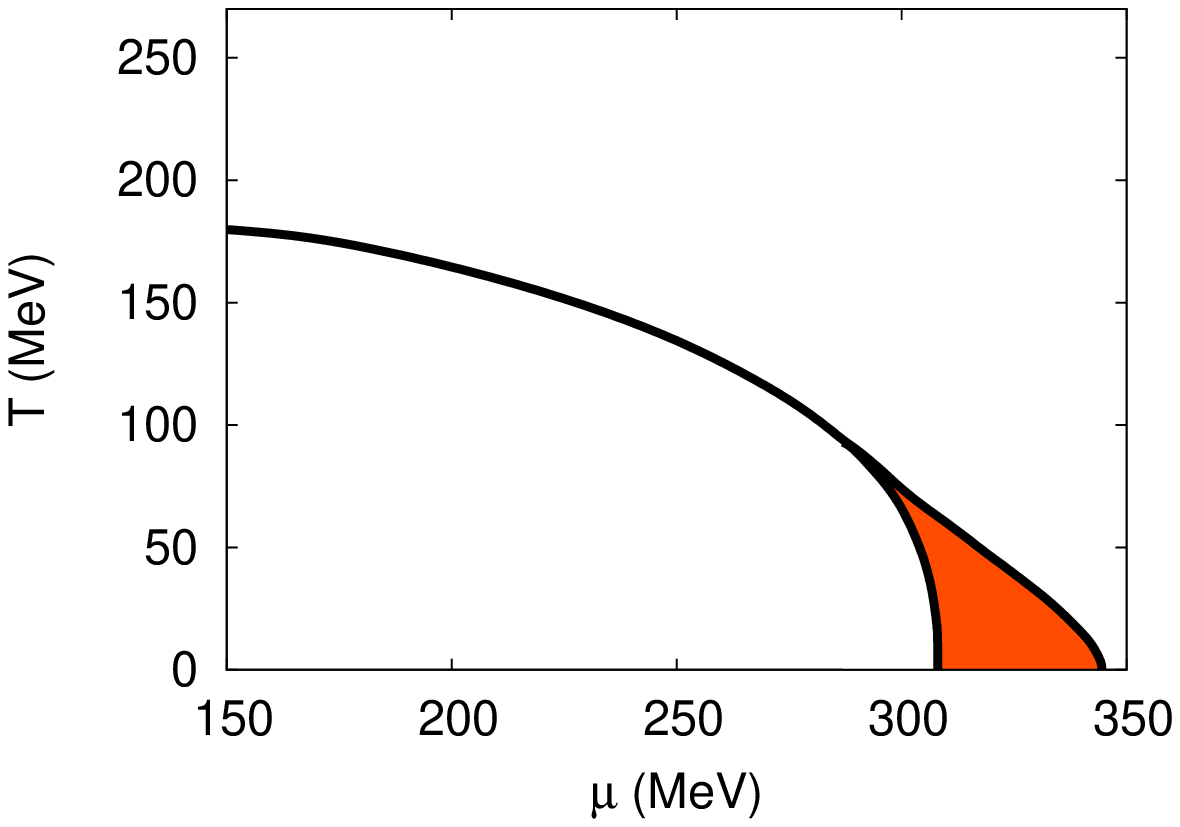}
\includegraphics[width=.32\textwidth]{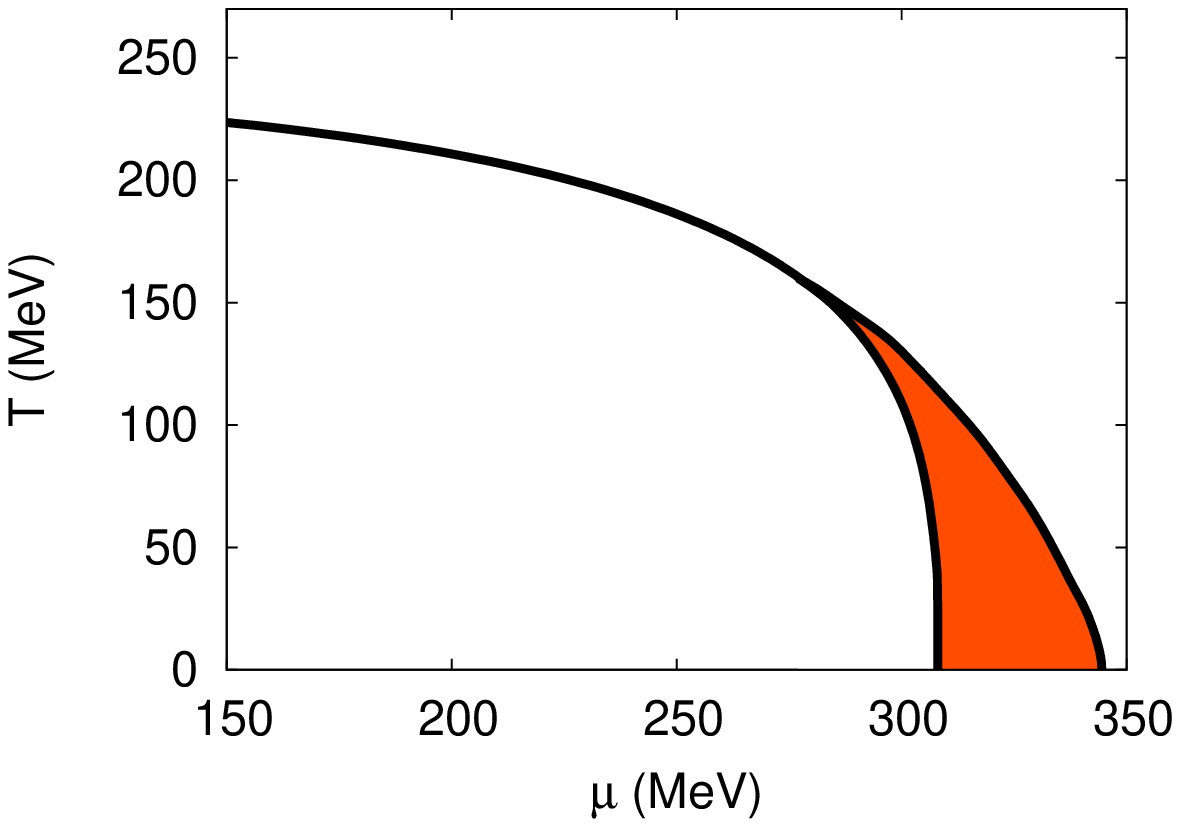}
\includegraphics[width=.32\textwidth]{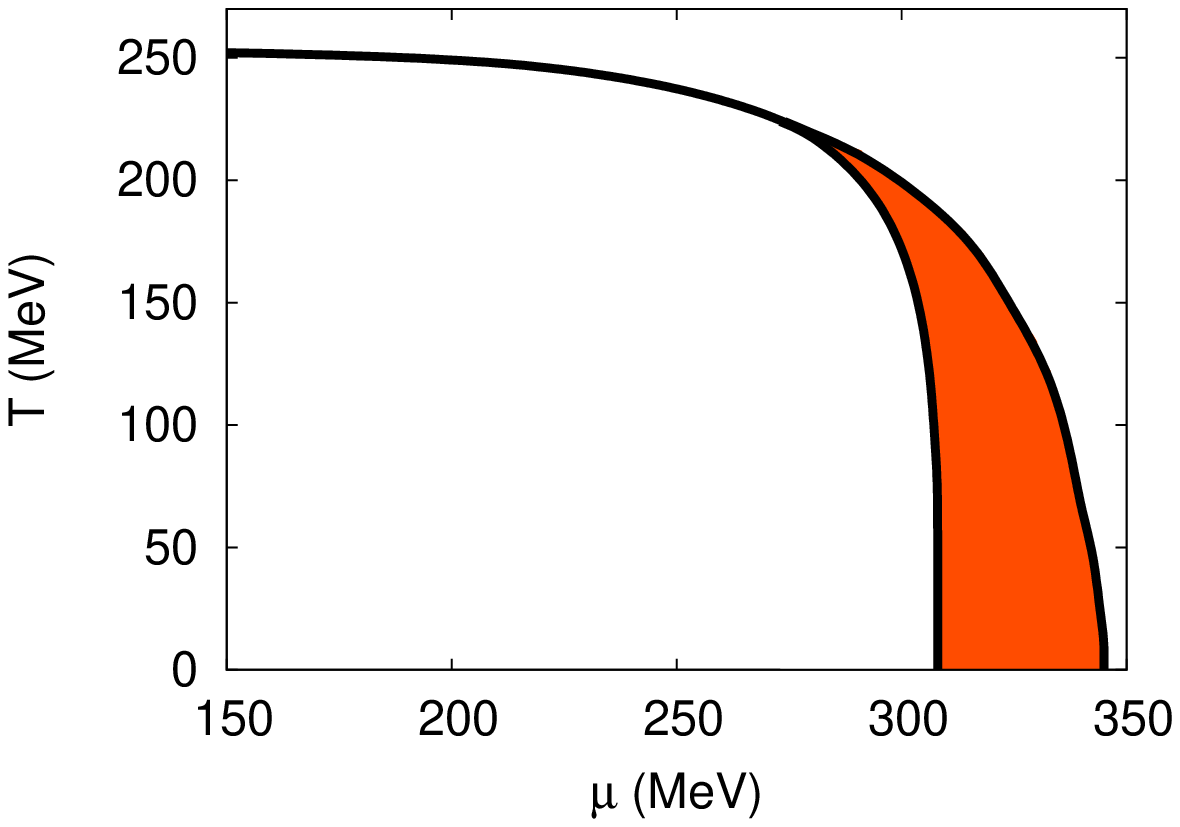}
\end{center}
\caption{PNJL phase diagram varying the number of colors: 
$N_c=3$ (left), $N_c=10$ (center), $N_c=50$ (right).
The shaded region indicates the inhomogeneous phase.}
\label{fig:PNJL}
\end{figure}

As the exact implementation of the Polyakov-loop expectation value
is not unique at arbitrary $N_c$, we follow Ref.~\cite{McLerran:2008ua} 
and change the function $f_\mathit{med}$, Eq.~(\ref{eq:fmed}),
in the thermodynamic potential into
 
\begin{eqnarray}
f_{med} &=& 
\theta(E_p-\mu)
\ell\, T \left( e^{-(E_p-\mu)/T} + e^{-(E_p+\mu)/T} \right) 
\nonumber
\\
&+& \theta(\mu-E_p)\left[ (\mu-E_p)
{}+ \ell\,T \left( e^{-(\mu-E_p)/T} +
e^{-(\mu+E_p)/T} \right) \right] \,,
\end{eqnarray}
which basically amounts to a leading-order expansion for small $\ell$ and 
is as such expected to be accurate at large $N_c$ in the confined phase. 
For the shape of the inhomogeneous modulations we consider the
one-dimensional solitonic ansatz, Eq.~(\ref{eq:msolitons}). 

Our results for $N_c =$ 3, 10 and 50 are shown in Fig.~\ref{fig:PNJL}.
By increasing the number of colors, the inhomogeneous phase is enlarged and 
stretches towards higher temperatures, approaching the upper limit given by 
the pure glue transition temperature ($T_c=270$ MeV in our parametrization of 
the Polyakov loop potential).
The transition lines between homogeneous and inhomogeneous phases
become more and more vertical, assuming a shape resembling the expected form 
for the phase diagram in the large $N_c$ limit \cite{McLerran:2007qj}. 
The size of the inhomogeneous phase is not dramatically enhanced in the $\mu$ direction.
In fact, in the PNJL model the Polyakov loop decouples from the NJL sector
at $T=0$. Therefore, since the latter is $N_c$-independent in mean-field
approximation, the transitions at $T=0$ are unchanged.

 \section{The inhomogeneous continent}

\begin{figure}
\begin{center}
\includegraphics[width=.45\textwidth]{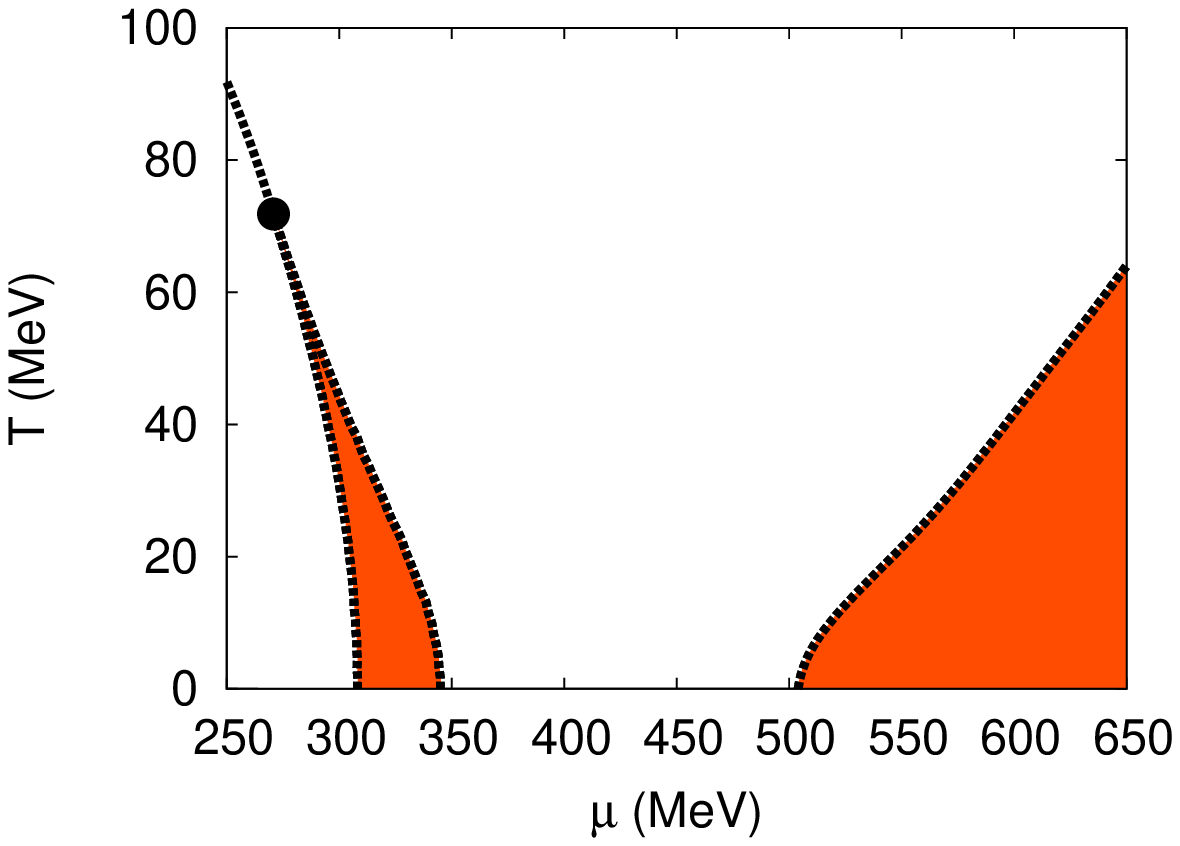}
\includegraphics[width=.45\textwidth]{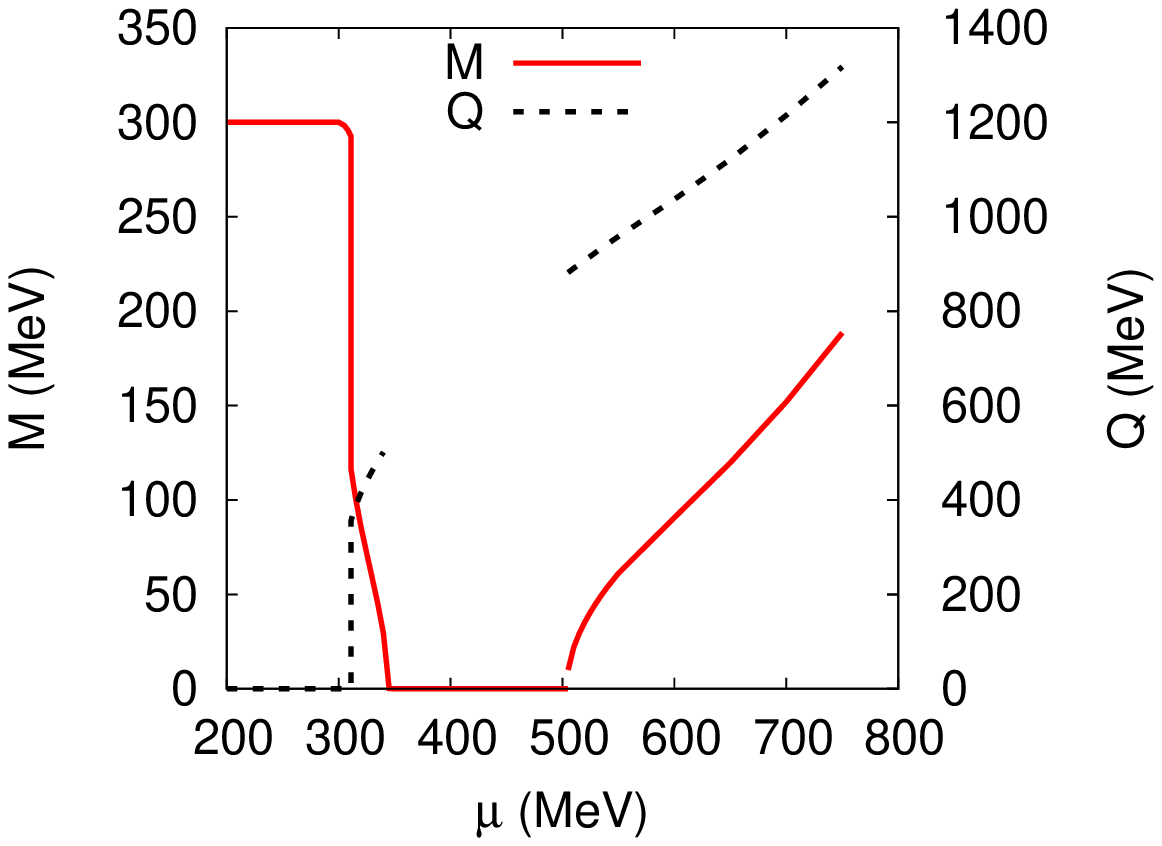}
\end{center}
\caption{
Left: Phase diagram with inhomogeneous island and continent.
Right: Amplitude and wave number for the CDW ansatz at $T=0$
using Pauli-Villars regularization.
}
\label{fig:continent}
\end{figure}

An unexpected feature emerges, when we extend our studies of the phase
diagram to higher values of $\mu$.
As shown in  Fig.~\ref{fig:continent}, above some critical value of the
chemical potential a second inhomogeneous phase appears, which seems to 
extend to arbitrarily high $\mu$ with steadily growing amplitude and 
wave number of the spatial modulation.  
This inhomogeneous ``continent'' is present for all kinds of spatial 
modulations we considered, and its onset is again given by a second-order 
phase transition.
In this case, GL analyses reveal that the phase boundary is the same for
all inhomogeneous solutions~\cite{Nickel:2009prl}.
Within some parametrizations, especially when one chooses a stronger coupling
to enforce a larger value of the constituent quark mass in vacuum,
the continent turns out to be even directly connected to the 
inhomogeneous island discussed until now.
In fact, indications for this behavior are already visible in 
Ref.~\cite{Nickel:2009}.

Since the continent appears at relatively high chemical potentials
and with large wave numbers, the first natural 
interpretation for it is that it is an artifact of the regularization. 
While this might indeed be the case, we will argue that the effect is
at least non-trivial. 
In order to achieve some better understanding on the observed behavior,
it might be worth analyzing in detail the mechanisms that within the model 
lead to the formation of an inhomogeneous condensate. 
Since, as argued above, we expect these considerations to be roughly 
independent of the particular ansatz chosen for the spatial modulation,
we focus on the simplest possible case, namely the CDW (\ref{eq:spiral}).

In this case, we obtain for the thermodynamic potential
\begin{equation}
\Omega - \Omega_\mathit{rest} 
= -N_fN_c\int_0^\infty \hspace{-3mm}
dE\, [\rho(E,M,Q)- \rho_\mathit{rest}(E)]
[f_\mathit{vac}(E,\Lambda) + f_\mathit{med}(E)] + \frac{M^2}{4G}, 
\label{eq:omega_minus_omegarest}
\end{equation}
where the integral corresponds to the kinetic term $\Omega_\mathit{kin}$,
Eq.~(\ref{eq:omegarho}),
and the $M^2$-term to the condensate term $\Omega_\mathit{cond}$,
Eq.~(\ref{eq:Omegacond}).
We have subtracted the free energy of the restored phase 
and explicitly indicated the dependence of the different terms
on the Pauli-Villars cutoff $\Lambda$, the amplitude $M$ and the wave number $Q$
of the CDW. 
In particular we note that only the function $f_\mathit{vac}$ depends
on the regularization whereas only the density of states depends on
the wave number.

\begin{figure}
\begin{center}
\includegraphics[width=.45\textwidth]{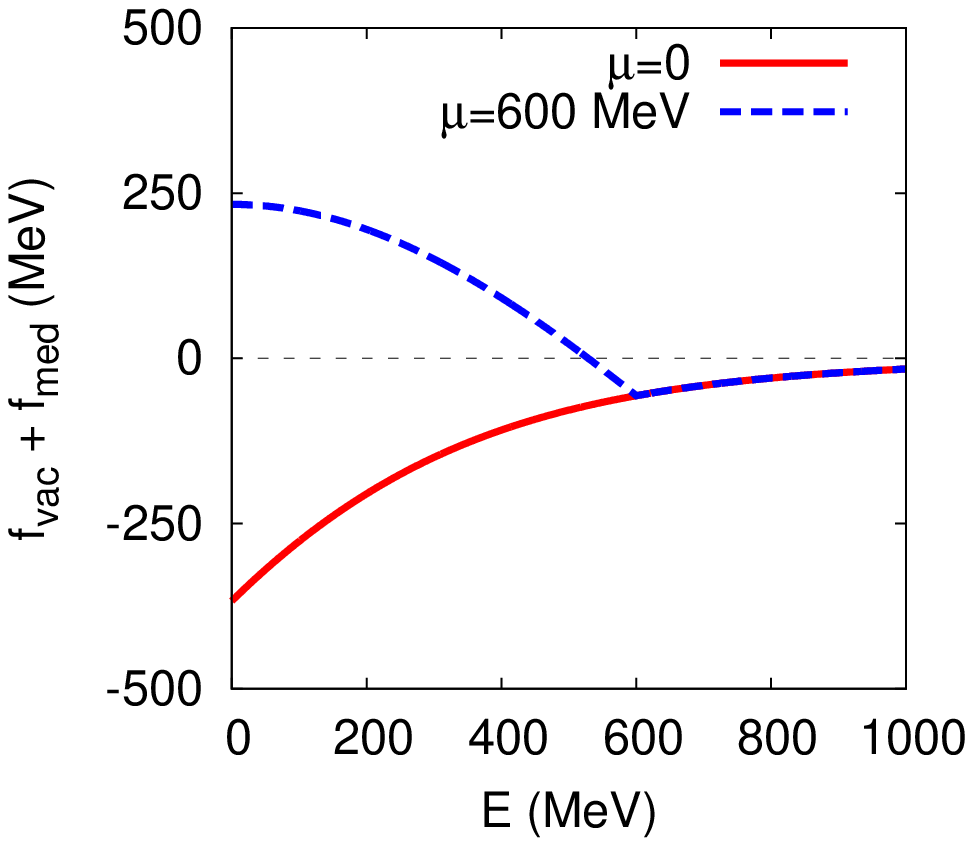}
\includegraphics[width=.45\textwidth]{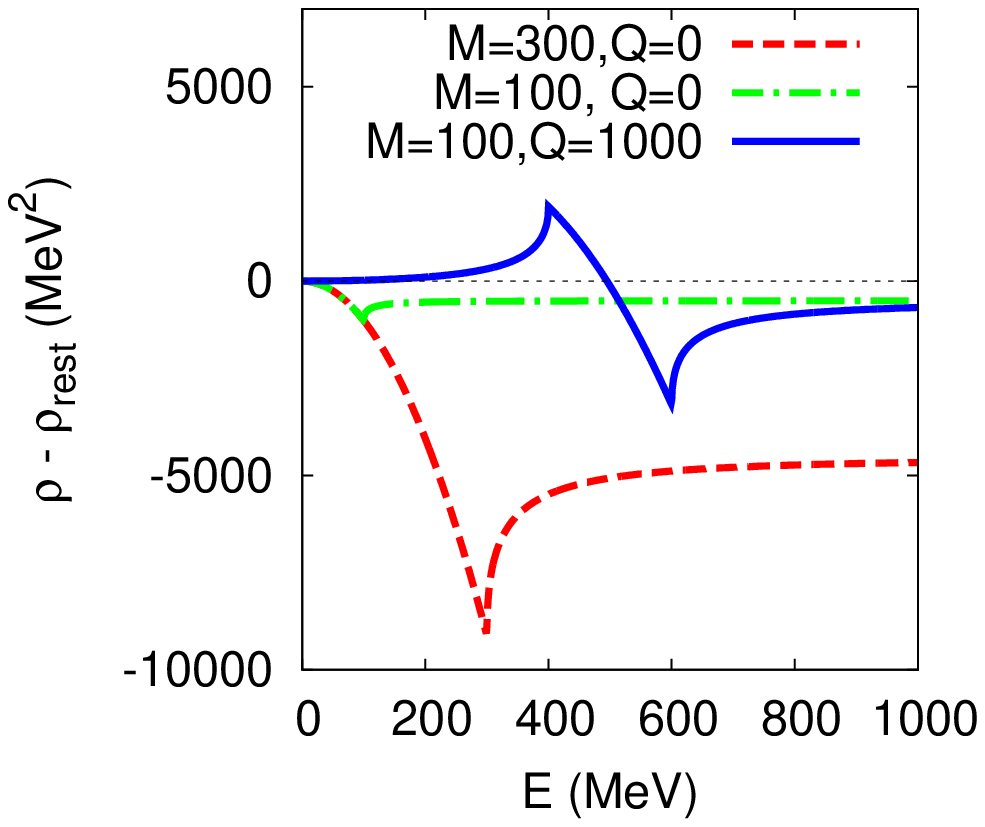}
\end{center}
\caption{Functions entering the integrand in the thermodynamic 
potential, Eq.~(\ref{eq:omega_minus_omegarest}).
Left: The red solid line represents the Pauli-Villars regularized function  
$f_\mathit{vac}$, corresponding to the  Dirac sea contribution,
while the blue dashed line indicates the sum $f_\mathit{vac}+f_\mathit{med}$
at $T=0$ and $\mu = 600$~MeV.
The two lines coincide above $E=\mu$, since the medium does not contribute 
beyond that point.
Right: Density of states $\rho(E,M,Q)$ relative to the 
restored case, $\rho_\mathit{rest}(E)\equiv\rho(E,0,0)$, for
homogeneous condensates with  
$M=300$~MeV (red dashed line), $M=100$~MeV (green dash-dotted 
line), and for a CDW with $M=100$~MeV, $Q=1000$~MeV (blue solid line). 
}
\label{fig:omegaintegrand}
\end{figure}

Since the condensate term always disfavors (homogeneous or inhomogeneous) 
chiral symmetry breaking, a necessary condition for a non-trivial
phase is that the first term is negative, \ie the integral must be
positive.
The functions which enter the integrand are displayed in 
Fig.~\ref{fig:omegaintegrand} for several cases.
As shown in the left panel, the Pauli-Villars regularized vacuum function
$f_\mathit{vac}$ (red solid line) is negative for all energies.
Hence, in vacuum, the density of states minus its value for the chirally 
restored solution should be negative as well to obtain a positive
integrand.
Indeed, as seen on the right, for $Q=0$ the difference 
$\rho - \rho_\mathit{rest}$ is always negative and its absolute value increases
with increasing $M$.
Thus, a larger constituent quark mass leads to an increase of the 
free-energy gain in the kinetic term, which is stabilized by 
the condensate term at some optimum value of $M$.
This is the usual mechanism for spontaneous chiral symmetry breaking in 
vacuum.

In contrast to $f_\mathit{vac}$, the function $f_\mathit{med}$ is positive.
Hence, when medium effects are included, the sum $f_\mathit{vac}+f_\mathit{med}$
becomes less and less negative at small energies until it eventually changes 
sign. 
As an example the case $T=0$, $\mu = 600$~MeV is shown in the left panel of 
Fig.~\ref{fig:omegaintegrand} (blue dashed line).
Since at $T=0$ the medium only contributes to energies $E<\mu$, the sum
$f_\mathit{vac}+f_\mathit{med}$ remains negative at large energies, 
but altogether the formation of homogeneous condensates, related to a 
negative function $\rho - \rho_\mathit{rest}$, becomes progressively 
disfavored by the medium contributions, 
thus leading to chiral restoration if we restrict ourselves to $Q=0$.

The situation changes, however, when we allow for a CDW with $Q> 2M$. 
In this case $\rho - \rho_\mathit{rest}$ is {\it positive} at small energies
and changes sign at $E \approx Q/2 - M^2/(2Q)$ (see 
Fig.~\ref{fig:omegaintegrand}, blue solid line on the right).
Hence, by properly choosing $Q$, it can be achieved that the factors 
$f_\mathit{vac}+f_\mathit{med}$ and $\rho - \rho_\mathit{rest}$ have the
same sign for all energies, thus more or less optimizing the free energy
gain in the kinetic part of the thermodynamic potential. 
As $f_\mathit{vac}+f_\mathit{med}$ changes sign slightly below $E=\mu$ 
at large $\mu$, this estimate yields $Q \approx 2\mu$ for the
favored value.

These arguments support the statement that the formation of an 
inhomogeneous chiral condensate is a medium-induced 
effect~\cite{Kojo:2009ha,Sadzikowski:2000,NT:2004}.
In our case it is technically related to the fact that both, 
$\rho - \rho_\mathit{rest}$ with $Q>2M$ and $f_\mathit{med}$, 
are positive at small energies. 
Therefore, since none of these functions but only the vacuum term
is affected by the regularization, 
the appearance of the continent does not immediately look like
a regularization artifact.

Of course, the dynamical formation of the inhomogeneous condensate
eventually depends on the interplay of $f_\mathit{vac}$ and $f_\mathit{med}$,
and the explicit expression for the Dirac sea term will naturally influence 
the quantitative details. 
For example, by allowing for a larger cutoff $\Lambda$, the vacuum part 
becomes larger and the second continent is shifted to higher 
chemical potentials. However, the same is true for the standard chiral
phase transition between homogeneous phases, where this is usually not
considered to be a major problem. 

The existence of the inhomogeneous continent is not restricted to
Pauli-Villars regularization, but the very same behavior is also present
when the vacuum term is regularized following the Schwinger proper-time 
prescription.
On the other hand, when the thermodynamic potential is regularized by
introducing a three-momentum cutoff,
it is clear from Eq.~(\ref{eq:mfH}), that a restriction of the in- 
and outgoing momenta also restricts the size of the wave number to which
they can couple. 
Hence, large values of $Q$ are strongly disfavored and 
the second inhomogeneous phase, if it exists at all, cannot extend to
arbitrary large values of $\mu$.
However, unlike the previous examples, this is {\it obviously} a 
regularization effect, since the suppression of large $Q$ is directly
caused by the cutoff.
In fact, for this reason the use of a three-momentum cutoff was 
abandoned in Ref.~\cite{NB:2009}
in the context of inhomogeneous color superconductors.\footnote{
For the same reason the authors of Ref.~\cite{Kojo:2011} introduce
a form factor, which restricts the quark momenta to the vicinity of
the Fermi surface, rather than to small absolute values.}

An example where large values of $Q$ are suppressed in a seemingly more 
physical way is the quark-meson (QM) model, defined by the Lagrangian 

\begin{equation}
 \mathcal{L}_{QM} = \bar\psi (i\gamma^\mu\partial_\mu - g(\sigma + i\gamma^5\tau_a \pi^a))\psi -\frac{1}{2}(\partial_\mu\sigma\partial^\mu\sigma
 + \partial_\mu\pi^a\partial^\mu\pi_a) 
 + \mathcal{U}(\sigma,\pi) \,.
 \label{eq:LQM}
 \end{equation}
Here $\sigma$ and $\vec \pi$ are elementary meson fields and 
$\mathcal{U}$ is a Mexican-hat type meson potential, leading to spontaneous 
chiral symmetry breaking.
While NJL and QM model are formally quite similar, in the latter the vacuum 
contribution is usually omitted when performing a mean-field treatment. 
On the other hand, for CDW ansatz, which for the QM model reads 
$\sigma(\vec x) = f_\pi \cos(\vec Q\cdot \vec x)$,
$\pi^3(\vec x) = f_\pi \sin(\vec Q\cdot \vec x)$,
the kinetic term of the mesons yields a contribution
\begin{equation}
    \Omega_\mathit{kin}^\mathit{mesons} = \frac{1}{2} f_\pi^2 Q^2\,, 
\label{eq:Omegakinmesons}
\end{equation}
which suppresses large values of $Q$
(see Ref.~\cite{Broniowski:2011} for more details). 
As a result, the inhomogeneous continent is not present in the QM model,
as far as we can tell numerically. 

From this one might naively conclude that the emergence of the continent
in the NJL model is due to absence of derivative terms in the Lagrangian. 
However, it is well known that in the NJL model the contribution 
(\ref{eq:Omegakinmesons}) is already contained in the vacuum part of the 
thermodynamic potential, which carries a dependence on $Q$ through
the energy spectrum~\cite{Broniowski:2011,NT:2004,NT:2004b}.
Indeed, if one expands
\begin{equation}
    \Omega_\mathit{vac}(M,Q) = \Omega_\mathit{vac}(M,0)
    + \beta_\mathit{vac,2}\,Q^2 + \beta_\mathit{vac,4}\,Q^4 + \dots
\label{eq:Omegaexpand}
\end{equation}
one can show in a regularization independent way that the so-called
spin stiffness is given by  
\begin{equation}
\beta_\mathit{vac,2} = \frac{1}{2}f_\pi^2 \,,
\label{eq:q2term}
\end{equation}
giving rise to a term like Eq.~(\ref{eq:Omegakinmesons})
in the thermodynamic potential.
Hence, if we were allowed to neglect all other terms in 
Eq.~(\ref{eq:Omegaexpand}), we could conclude that large values of
$Q$ and, thus, the inhomogeneous continent are suppressed in the same way 
as in the QM model.
However, in the continent region, $Q$ is certainly not small 
(compared with $M$ or $f_\pi$). 
In any case, it is clear that the stability against large values of 
$Q$ cannot be discussed in terms of a Taylor expansion for small $Q$.

We therefore look at the higher orders in the series in order to get an
idea how these additional contributions (which become more and more relevant 
at higher values of $Q$) influence the previous considerations.
In particular the coefficient of the $Q^4$-term is dimensionless and 
therefore expected to stay finite, even if the cutoff is sent to infinity.
For instance, if we employ proper-time regularization
(see \eg Ref.~\cite{NT:2004}), we obtain
\begin{equation}
  \beta_\mathit{vac,4}
  \,=\, - 2 N_f N_c\,e^{-M^2/\Lambda^2} 
  \frac{\Lambda^2 - M^2}{192\Lambda^2\pi^2} 
  \;=\; -\frac{N_f N_c}{81\pi^2} + {\cal O}(\frac{M^2}{\Lambda^2})
  \,.
  \label{eq:q4term}
\end{equation}
The most important observation is that this term is negative, \ie
it weakens the effect of the $Q^2$-term.
Although the value of the coefficient is relatively small, 
the $Q^4$-term dominates the $Q^2$-term when 
$Q$ is of the order of 10 times $f_\pi$.

\begin{figure}
\begin{center}
\includegraphics[width=.5\textwidth]{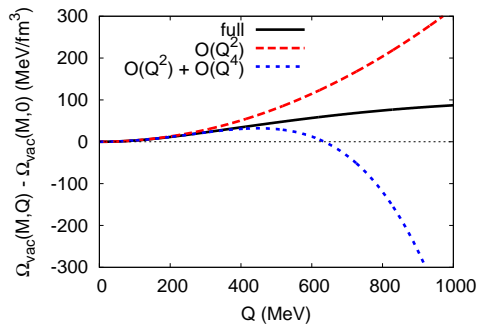}
\end{center}
\caption{Proper-time regularized vacuum thermodynamic potential for a 
CDW with fixed amplitude $M=100$~MeV as a function of the wave number $Q$, 
using parameters from \cite{NT:2004}.
The value for $Q=0$ has been subtracted. The full result (black solid line)
is compared with the Taylor series truncated at order $Q^2$ (red dashed line)
and order $Q^4$ (blue dotted line).
}
\label{fig:OmegaTaylor}
\end{figure}

In Fig.~\ref{fig:OmegaTaylor} we compare the full 
proper-time regularized vacuum thermodynamic potential with the
results of a Taylor expansion to the orders $Q^2$ and $Q^4$. 
In agreement with our considerations above, the $Q^2$-result gets
reduced by the higher orders. 
Although the effect is overestimated by the $Q^4$-term, this leads
to a much more moderate increase of $\Omega_\mathit{vac}$ in the 
continent region. As a consequence, the vacuum contributions do 
not disfavor the formation of a CDW strongly enough in this region
to compete with the favoring effects of the medium contributions.

A very systematic investigation of the large-$Q$ behavior of the
vacuum effective potential has been performed more than 20 years
ago in Ref.~\cite{Broniowski:1990gb}. 
Although it turned out that the exact behavior beyond the 
universal $Q^2$-order is strongly regularization dependent,
in all cases considered, $\Omega_\mathit{vac}$ eventually becomes 
negative, meaning that even the vacuum is unstable against the formation
of a CDW with very large $Q$.
While this is clearly unphysical and should therefore be ignored,
there is no reason why corrections to the $Q^2$-terms should
not be present at all. 
Unfortunately, these corrections are very model dependent and
theoretical input from outside is needed to pin them down.
In fact, inhomogeneous phases at arbitrary large $\mu$ have been 
predicted for the $1+1$-dimensional Gross-Neveu model~\cite{Thies:2006ti}
as well as for QCD in the large-$N_c$ 
limit~\cite{Deryagin:1992}\footnote{For $N_c=3$, they are, however,
disfavored against color superconductivity~\cite{Shuster:1999}.}.
Therefore, the continent may be not as exotic as it appears.

\section{Conclusions}

We presented some of our recent results related to inhomogeneous 
chiral symmetry breaking phases in the NJL model.
Thereby one focus was a comparison of different spatial modulations 
of the chiral condensate, including an ``egg carton''-like two dimensional
ansatz and several one-dimensional functions. 
For all of them we observed the emergence of an inhomogeneous ``island'' 
around the region where the usual first-order chiral phase transition
would occur for homogeneous phases.

A comparison of the free energies at $T=0$
seems to support the idea that one-dimensional 
modulations are favored over higher dimensional ones.
In particular, solitonic solutions~\cite{Nickel:2009}
inspired by analytical studies of 1+1-dimensional models 
seem to constitute the favored state. 
While only one kind of two-dimensional 
ansatz has been considered so far, it may very well be that this feature 
is valid in general for all kinds of 2D modulations.
In fact, it has already been argued in a Ginzburg-Landau study
that this is the case in proximity of the critical point~\cite{Abuki:2011}. 

The inhomogeneous island also exists in the Polyakov-loop extended
NJL model. In this framework we have studied the effects of varying
the number of colors on the phase diagram.
We found that with increasing $N_c$ the inhomogeneous phase is 
stretched towards higher temperatures so that the boundaries
between homogeneous and inhomogeneous phases become more and more
vertical. On the other hand, the size of the inhomogeneous phase in 
the $\mu$ direction is not enhanced dramatically.

Finally, we investigated the origin of the inhomogeneous ``continent'',
\ie a second inhomogeneous phase which appears in our calculations at
higher chemical potential and does not seem to end. 
We discussed the reliability of our model in the region where the 
continent appears, focusing on effects of the regularization process.
It turned out that the continent is not the result of a {\it trivial}
cutoff artifact but, unfortunately, regularization dependencies, 
including a known vacuum instability~\cite{Broniowski:1990gb}, 
preclude the model from giving definite answers about the phase structure 
in this regime.

\bigskip

We thank the organizers for a very interesting workshop. 
Travel support by HIC for FAIR (S.C.) and by 
the DFG under contract BU 2406/1-1 (M.B.) is gratefully acknowledged.
This work was partially supported by
the Helmholtz Alliance EMMI, 
the Helmholtz International Center for FAIR, 
and by the Helmholtz Research School for Quark Matter Studies H-QM.

\end{document}